\PassOptionsToPackage{table}{xcolor} 
\documentclass[11pt]{article} 
\PassOptionsToPackage{numbers}{natbib}
\usepackage{arxiv}
\usepackage{natbib}

\usepackage{times}
\usepackage{latexsym}
\usepackage[T1]{fontenc}
\usepackage[utf8]{inputenc}
\usepackage{microtype}
\usepackage{inconsolata}
\usepackage{graphicx}
\usepackage{wrapfig}

\usepackage{amsmath,amsfonts,bm}

\def\eqref#1{equation~\ref{#1}}

\def\1{\bm{1}}

\DeclareMathAlphabet{\mathsfit}{\encodingdefault}{\sfdefault}{m}{sl}
\SetMathAlphabet{\mathsfit}{bold}{\encodingdefault}{\sfdefault}{bx}{n}

\usepackage{url}
\usepackage{amsmath}
\usepackage{amsfonts}
\usepackage{nicefrac}
\usepackage{xcolor}
\usepackage{booktabs}
\usepackage{multirow}
\usepackage{makecell}
\usepackage{adjustbox}
\usepackage{subcaption}
\usepackage{tabularx}
\usepackage{algorithm}
\usepackage{algpseudocode}
\usepackage{tcolorbox}
\usepackage{listings}
\usepackage{newfloat}
\usepackage{float}
\usepackage{placeins}
\usepackage{needspace}
\usepackage{microtype}
\usepackage[capitalize]{cleveref}

\lstdefinestyle{promptbox}{
  basicstyle=\ttfamily\footnotesize,
  frame=single,
  framesep=8pt,
  rulecolor=\color{black},
  framerule=0.8pt,
  backgroundcolor=\color{white},
  numbers=none,
  showstringspaces=false,
  showspaces=false,
  showtabs=false,
  breaklines=true,
  breakatwhitespace=true,
  tabsize=2,
  aboveskip=1.2em,
  belowskip=1.2em,
  xleftmargin=8.8pt,
  xrightmargin=8.8pt,
  columns=fullflexible,
  breakindent=0pt,
  keepspaces=true,
  gobble=0
}

\DeclareFloatingEnvironment[
    fileext=lop,
    listname={List of Guides},
    name=Guide
]{guide}
\crefname{guide}{Guide}{Guides}
\Crefname{guide}{Guide}{Guides}
\tcbset{
  colback=white,
  colframe=cyan!90!blue,
  arc=3mm,
  boxrule=0.8pt,
  left=6pt,
  right=6pt,
  top=6pt,
  bottom=6pt,
  toptitle=1pt,
  bottomtitle=1pt,
  lefttitle=6pt,
  fontupper=\fontsize{6.5}{8.5}\selectfont\ttfamily\raggedright
}
\definecolor{lightorange}{RGB}{255,218,170}

\title{Federation over Harness}
\author{Dixi Yao\footnotemark[0] \thanks{Correspondence
to: Dixi Yao $<$dixi@uchicago.edu$>$} \quad
Tahseen Rabbani \quad
Tian Li \\University of Chicago 
}

\begin{document}
\maketitle
\renewcommand{\topfraction}{0.85}
\renewcommand{\dbltopfraction}{0.85}
\renewcommand{\textfraction}{0.1}
\renewcommand{\floatpagefraction}{0.8}
\renewcommand{\dblfloatpagefraction}{0.8}
\setcounter{topnumber}{3}
\setcounter{dbltopnumber}{3}

\begin{abstract}

Modern LLM-powered agents increasingly improve through experience sharing, where distributed agents distill local reasoning experience into reusable knowledge, such as a shared insight library. Existing approaches, however, still heavily rely on manually designed prompts and workflows that determine both how local agents extract useful reasoning traces and how a central/global agent aggregates them to obtain common patterns and insights. We propose Federation over Harness (FoH), a framework that automatically evolves the \textit{harness} for both local and global components in this distributed multi-agent reasoning pipeline. A server agent evolves the client harness for task solving and self-reflection while, at the same time, evolving its own harness for aggregating distributed reflections, maintaining shared knowledge, and updating subsequent client harnesses. Importantly, FoH requires no extra validation signals, such as ground-truth labels or validation data, during evolution; it learns solely from self-generated reflections produced by distributed agents. 
Experiments on multiple benchmarks show that FoH consistently improves over single-agent baselines with a fixed harness and with harness self-evolution by 53.57\% and 29.76\%.
\end{abstract}

\section{Introduction}

The capabilities of modern LLM-powered agents increasingly depend not only on the underlying model, but also on the \textit{harness}, which governs tools, memory, feedback, and execution workflows. Frontier systems such as Claude Code~\citep{anthropic2025claudecode}, DeepSeek Harness~\citep{deepseek2026harness}, and Hermes Agent~\citep{nousresearch2026hermesevolution} therefore emphasize harness design. Recent work further studies \textit{harness evolution}, where execution experience is provided to a meta-agent to improve the harness~\citep{lee2026recursive,lee2026meta,zhang2026self,zhang2026harnesscompass,yin2025godel,zhang2026hyperagents}.

Meanwhile, distributed multi-agent systems increasingly improve through experience sharing, where independently operating agents distill local reasoning experience into reusable knowledge in a federated manner, because the runtime environments of agents vary widely and are hard to centralize. Each agent may use a different operating system, file system, database, compiler, and so on. FoT~\citep{yao2026federation}, for example, aggregates agent self-reflections into a shared insight library serving as collective memory for future tasks. However, existing federated systems rely on manually designed client and server prompts and workflows.

As harnesses grow more complex, manual design becomes increasingly restrictive or requires extraordinary human effort. We therefore ask: \textit{can harnesses that govern both the local and global components of a distributed multi-agent system be evolved automatically?} To this end, we propose \emph{Federation over Harness (FoH)}, a framework for federated harness evolution driven solely by agent-generated reflections. Humans provide only a lightweight, plaintext \textit{guide} describing the allowable harness search space and expected behavior.

In FoH, clients solve local tasks and send reflections to a central server, which first evolves its own server harness for processing distributed client agent reflections and then uses the updated harness to evolve the shared client harness. The shared insight library is treated as part of the harness and evolves jointly with prompts, workflows, and other components. Importantly, FoH requires no label signals during evolution, such as ground-truth labels or scores. Benchmark evaluation results are inaccessible to the agents during evolution. This is particularly important in federated settings, where agents may only access private or unlabeled local tasks, making centralized validation impractical~\citep{pan2026evolving}. Moreover, FoH allows the server agent to operate in different environments from the client agents, since they may have different tools (excluding those evolved by the harness-design agents), systems, and other resources.

In summary, FoH extends federated harness evolution beyond shared reflections by enabling both client and server harnesses to evolve through reflection-only optimization. Our contributions are summarized as follows.
(1) We propose \textit{Federation over Harness (FoH)}, 
a federated framework that uses distributed client reflections to evolve both a shared client harness and the server harness responsible for updating it. Within harness search spaces specified by lightweight textual guides, FoH makes both task execution and the harness-evolution procedure adaptable, without external validation or labels (\Cref{sec:method}). (2) We comprehensively evaluate FoH on held-out real-world tasks (ClawEval~\citep{ye2026claw}) and data-driven scientific discovery tasks (ScienceAgentBench~\citep{chen2025scienceagentbench}). FoH outperforms the SOTA single-agent harness by 53.57\%, non-federated harness evolution by 29.76\%, and fixed-harness federated self-evolution by 22.38\%, while reducing task-completion tokens by 34.09\%. Evolving the server harness further improves performance by 4.50\% over a fixed server harness (\Cref{sec:evaluation}).
(3) We conduct detailed case studies and component-level analyses to investigate how client and server harnesses evolve, how different harness components are utilized across tasks, and how their interactions contribute to performance improvements (\Cref{sec:evaluation}).

\section{Related Work}
\textbf{Reasoning with Self-Reflection.} There is extensive prior work on (iterative) prompt optimization~\citep{zhou2022large,fernando2023promptbreeder,zelikman2023self}, especially work that uses LLM-generated feedback to iteratively refine task prompts~\citep{yuksekgonul2024textgrad,wang2024promptagent,tong2025evoprompt}.
Recent work
extends this idea to agentic workflows, where agents reflect on their own execution trajectories without an external reward model or feedback and distill reusable skills into the prompt~\citep{didolkar2025metacognitive,yang2024buffer,wang2024agent}. These works motivate our use of self-generated reflections as the signal for improving an agentic system. FoH builds on this paradigm but uses the resulting traces not only to improve task solving but also to drive client and server harness evolution.

\textbf{Harness Evolution.} Modern agentic systems place increasing emphasis on the harness alongside the LLM itself. Following the success of recursive self-improvement, prior work has explored self-evolving agent harnesses~\citep{lee2026recursive,lee2026meta,zhang2026self,zhang2026harnesscompass}. Recent work further proposes using editable programs to represent agent harnesses, allowing a meta-agent to modify the agent runtime~\citep{yin2025godel,zhang2026hyperagents,shao2026contextlanguagemodels}. These works recursively improve agent harnesses, but typically rely on extra validation data, human signals, or verifiers to identify good harnesses.
In contrast, FoH relies solely on self-reflection traces, without labels or judgment feedback from either clients or the server. The closest work, \citet{pan2026evolving} (RHO), also studies harness optimization without external validation data. However, RHO focuses on single-agent settings, and employs a fixed, manually designed evolution algorithm with predefined numbers of task rollouts and re-solving iterations where its evolution algorithm remains unchanged throughout optimization. 

\textbf{Federated Agents.} Recent work extends agent evolution from isolated agents to distributed experience sharing. Federation over Text (FoT)~\citep{yao2026federation} aggregates reasoning-derived information from multiple agents into a reusable insight library, while FederatedSkill~\citep{yang2026federatedskill} communicates and aggregates structured skill-library updates, without involving harness updates. EvolveNet~\citep{nie2026evolvenet} similarly collects reflections from federated agents but relies on an additional server-side validation set to select the aggregated harness. In this work, we focus on validation-free client and server harness evolution. Validation-free means that the whole system evolves without external task-performance supervision, but permits runtime checks within the harness, such as executing generated code to detect errors. 

\section{Federation over Harness (FoH)}
\label{sec:method}
We define an agent system as a language model together with its harness,
$\mathcal{A}=(\mathcal{M},\mathcal{H})$. FoH consists of a central server agent $\mathcal{A}_s$ and $N$ client agents
$\mathcal{A}_{c_1},\ldots,\mathcal{A}_{c_N}$. Clients independently solve
different but related user tasks using a shared client harness $\mathcal{H}_c$, while
the server uses a server harness $\mathcal{H}_s$ to aggregate knowledge across clients and improve the shared client harness (including updating the shared insight library).

\begin{figure*}[!htbp]
\centering

\begin{adjustbox}{valign=t}
\begin{minipage}{0.48\textwidth}
\begin{algorithm}[H]
\footnotesize
\caption{Federation over Harness}
\label{alg:foh}

\begin{algorithmic}[1]

\Require $N$ client agents $\{\mathcal{A}_{c_1}, \cdots,\mathcal{A}_{c_N}\}$ with tasks $\{T_1,\cdots,T_N\}$; server agent $\mathcal{A}_s$.
\Ensure Client harness guide $G_c$; server harness guide $G_s$; LLMs $M$; initial server harness $\mathcal{H}_s^{(0)}$.

\Statex \textbf{Server initialization}

\State $\mathcal{H}_s^{(1)}
    \gets \mathcal{A}_s(M,\mathcal{H}_s^{(0)};G_s)$
\State $\mathcal{H}_c^{(1)}
    \gets \mathcal{A}_s(M,\mathcal{H}_s^{(1)};G_c)$

\Statex \textbf{Parallel client execution at iteration $t$}
\ForAll{clients $i=1,\ldots,N$ \textbf{in parallel}}
    \State Pull the latest client harness $\mathcal{H}_c^{(t)}$
    \State $(R_i^{(t)},\mathrm{Answer}_i^{(t)})
        \gets \mathcal{A}_{c_i}
        (M_i,\mathcal{H}_c^{(t)};T_i)$
    \State Return $\mathrm{Answer}_i^{(t)}$; upload $R_i^{(t)}$ to server
\EndFor

\Statex \textbf{Server-side evolution at iteration $t$}
\If{reflections $R_{1:N}^{(t)}$ have been collected}
    \State $\mathcal{H}_s^{(t+1)}
        \gets \mathcal{A}_s
        (M,\mathcal{H}_s^{(t)};
        G_s,R_{1:N}^{(t)},\mathcal{H}_s^{(t)},\mathcal{H}_c^{(t)})$
    \State $\mathcal{H}_c^{(t+1)}
        \gets \mathcal{A}_s
        (M,\mathcal{H}_s^{(t+1)};
        G_c,R_{1:N}^{(t)},\mathcal{H}_c^{(t)})$
\EndIf

\end{algorithmic}
\end{algorithm}
\end{minipage}
\end{adjustbox}
\hfill
\begin{adjustbox}{valign=t}
\begin{minipage}{0.49\textwidth}
\centering
\includegraphics[
    width=0.8\linewidth,
    trim=0 0 0 0,
    clip
]{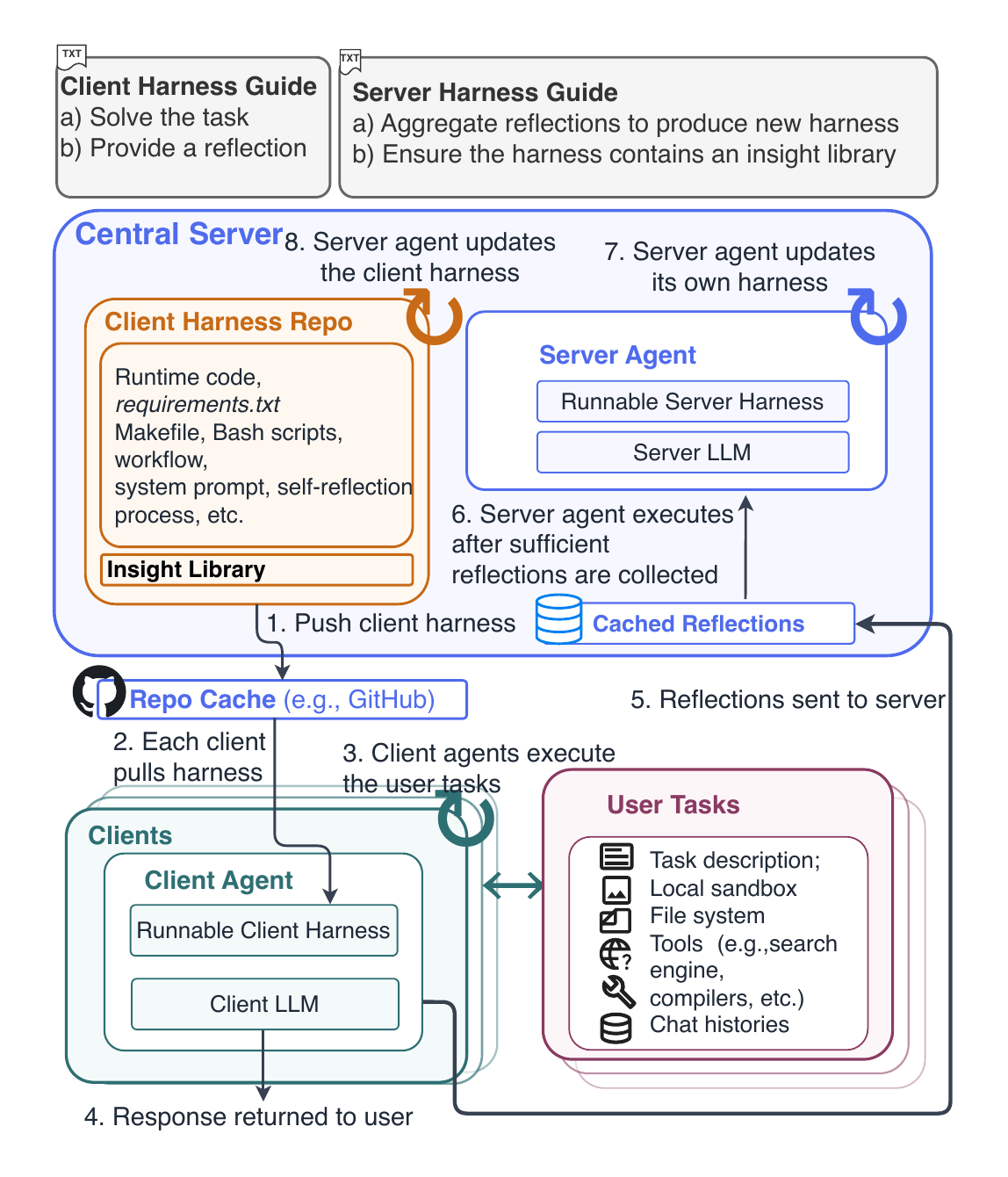}
\captionsetup{skip=2pt}
\captionof{figure}{Overview of the FoH workflow.}
\label{fig:workflow}
\end{minipage}
\end{adjustbox}

\end{figure*}

The server first initializes $\mathcal{H}_s$ and $\mathcal{H}_c$ from lightweight descriptions of their search spaces and purposes. In each iteration, clients execute local tasks with $\mathcal{H}_c$ and upload self-reflections; the server first updates its own harness with them and then executes under the updated server harness to produce the next client harness (\Cref{alg:foh} and \Cref{fig:workflow}).
We next describe the harness search space, client-side execution, server-side evolution, and the implementation.

\renewcommand{\algorithmicensure}{\textbf{Input:}}

\begin{table*}[!htbp]
\centering
\footnotesize
\captionsetup{skip=4pt}
\caption
{\textbf{Results on real-world daily tasks.} Token columns report output tokens; the maximum per LLM call is 16,384 (client) and 32,768 (server), with at most 30 LLM calls per task and no runtime limit. We run 15 iterations; isolated-agent performance is the best of 15 independent runs. Base LLM: DeepSeek-V4-Flash; judge: GPT-Luna-5.6-Pro (High). Sandbox: CPUs, 64\,GB memory, Python 3.10, and Conda; other settings follow benchmark defaults.}
\label{tab:foh_main}
\setlength{\tabcolsep}{4pt}
\renewcommand{\arraystretch}{0.91}
\begin{adjustbox}{max width=\linewidth}
\begin{tabular}{@{}llccccc@{}}
\toprule
\textbf{Type}
& \textbf{Method}
& \makecell{\textbf{Existing Tasks}\\\textbf{PinchBench V1}}
& \makecell{\textbf{Held-Out Tasks}\\\textbf{PinchBench V2 $\setminus$ V1}}
& \makecell{\textbf{New Tasks}\\\textbf{ClawEval}}
& \makecell{\textbf{Total}\\\textbf{Tokens}}
& \makecell{\textbf{Task Completion}\\\textbf{Tokens}} \\
\midrule
\multirow{3}{*}{Baseline}
& Isolated Agents (ReAct~\citep{yao2022react}) & 83.98 & 84.80 & 56.60 & 645K & 645K \\
& FoT~\citep{yao2026federation} & 92.77 & 90.53 & 74.41 & \textbf{453K} & 416K \\
& RHO~\citep{pan2026evolving} & 86.11 & 80.72 & 65.83 & 15M & 377K \\
\midrule
\multirow{4}{*}{FoH Ablations}
& FoH ($\mathcal{H}_s$ fixed at $\mathcal{H}_s^{(1)}$) & 93.10 & 91.51 & 76.53 & 465K & 424K \\
& FoH (no insight library) & 83.02 & 65.86 & 62.34 & 498K & 375K \\
& FoH (no local reflection) & 73.45 & 68.32 & 61.29 & 501K & \textbf{312K} \\
& FoH (no tool evolution) & 92.12 & 86.71 & 70.43 & 511K & 417K \\
\midrule
\rowcolor{lightorange}
\textbf{Ours} & \textbf{FoH} & \textbf{94.47} & \textbf{92.86} & \textbf{78.44} & 519K & 414K \\
\bottomrule
\end{tabular}
\end{adjustbox}
\end{table*}

\textbf{Harness Search Space and Guide.}
Existing agent systems such as Claude~\citep{anthropic2025claudecode}, Codex~\citep{openai2025codex}, and DeepSeek~\citep{deepseek2026harness} adopt different notions of an agent harness. We define the harness as the components that can be revised by agents. Harness $\mathcal{H}$ consists of three categories: an \emph{insight library}, \emph{prompt-level components}, and \emph{tool-level components}. The insight library, inherited from FoT~\citep{yao2026federation}, stores reusable knowledge and experience. Prompt-level components specify the contextual information provided to the language model together with the task input, including skills, system prompts, and tool descriptions. Tool-level components specify executable functionality invoked based on model outputs, including scripts, programs, and functions for operations such as database queries, web search, and conversation-history retrieval. In contrast, task descriptions, access permissions, conversation histories, and other user-owned resources remain fixed inputs from users. A harness must therefore be defined within a clear operational scope. Allowing unconstrained modifications, such as replacing system-level components (e.g., the OS kernel) unavailable in the target environment, may produce undeployable solutions. We introduce a lightweight plaintext \textit{guide} $G$ that specifies this scope by describing which components the harness can access and modify, with separate guides $G_c$ and $G_s$ for the client and server harnesses (\Cref{sec:guide}).

\textbf{Client Agents.}
The client guide $G_c$ requires each client harness to solve its local task and generate a reflection trace ($R_i$) from its own execution trajectory. The reflection has access only to the local task, response, and trajectory, with no ground-truth answer, reward, or held-out validation example. Each client receives the latest shared harness $\mathcal{H}_c$, integrates user-provided resources (tools, environments, file systems, and chat history), and returns the response to the user, uploading only $R_i$ to the server (\Cref{promptclientguide}).

\textbf{A Server Agent Managing Agent Evolution.}
Rather than relying on a manually specified harness-optimization procedure, FoH uses a server agent $\mathcal{A}_s$ and its harness $\mathcal{H}_s$ to optimize both the server and client harnesses. Operationally, the server harness governs how the server processes client reflections, maintains the shared insight library, and revises the client harness. 
In \Cref{alg:foh}, we consider a synchronous updating scenario and \Cref{sec:future} describes extensions to the asynchronous case. 
At each iteration, the server collects client reflections ($R_{1:N}^{(t)}$) and performs two sequential updates. First, it runs under $\mathcal{H}_s^{(t)}$ to produce $\mathcal{H}_s^{(t+1)}$, using the server guide, the collected reflections, and the current server and client harnesses. Second, it runs under the updated server harness $\mathcal{H}_s^{(t+1)}$ to produce $\mathcal{H}_c^{(t+1)}$, using the client guide, the same reflections, and the current client harness. 
The shared insight library is updated on the server side, evolves jointly with other components of the server harness, and is used on the client side as external knowledge. FoH therefore jointly evolves the shared knowledge and the algorithm used to aggregate and use it.

The only human-provided artifacts are $G_s$ and $G_c$, which define the server and client harness search spaces and their roles in the federation topology. Unlike RHO, which manually fixes workflow choices such as the number of rollouts per task and of candidate harnesses per evolution step, FoH lets the evolving harnesses determine these choices, including how many rounds of client reflection to perform and how the server generates, evaluates, and selects new harnesses.

\textbf{Implementation Details.}
We represent each harness as a repository containing the code required for execution. Clients obtain updated harnesses through repository synchronization (e.g., Git pull). At execution, the harness takes a user query, integrates available user-provided resources (e.g., tools, chat histories, and files), executes its workflow, and returns a response. The initial server harness $\mathcal{H}_s^{(0)}$ is instantiated as a Hermes Agent~\citep{nousresearch2026hermesevolution} using ReAct~\citep{yao2022react}. Our implementation is available at \url{https://github.com/dixiyao/FoH}.

\section{Evaluation}
\label{sec:evaluation}
\subsection{Settings and Baselines}
We evaluate whether FoH improves agent performance under a validation-free evolution protocol and isolate the contributions of federation, full-harness evolution, and the evolvable server optimizer. All methods use the same LLM and harness search space defined by identical guides. We compare against a fixed ReAct harness (Isolated Agents) and RHO~\citep{pan2026evolving}; for RHO, we retain its predefined optimization procedure while placing it in the same federated client/server setting. We further ablate FoH by: (1) freezing $\mathcal{H}_s^{(1)}$ while continuing to evolve $\mathcal{H}_c$; (2) removing the insight-library requirement from $G_s$; (3) removing the local reflection requirement from $G_c$ and uploading raw trajectories instead; (4) with initial $\mathcal{H}_s^{(1)}$ and $\mathcal{H}_c^{(1)}$, restricting evolution to prompts rather than the full harness. Thus (2) evolves only prompt- and tool-level components while (4) evolves only prompt-level components and the insight library. 

We also include FoT~\citep{yao2026federation}, in which clients upload reflections and the server aggregates them into an insight library without evolving prompts or other harness components, i.e., only the insight library evolves. All methods except the fixed Isolated Agents undergo the initialization in Lines 1-2 of \Cref{alg:foh} with the corresponding guides. Since the evolution is validation-free, we evolve every method for a fixed number of iterations and report performance on existing and held-out tasks using the final-iteration harness.

\subsection{Applications}
\label{sec:app}
\textbf{Real-World Daily Tasks: PinchBench and ClawEval.} We use the 23 tasks in PinchBench~\citep{pinchbench} V1 as existing tasks and the 124 tasks subsequently added in V2 as held-out tasks. Only the 23 V1 tasks are used to generate client reflections and evolve the harness; the held-out PinchBench tasks and ClawEval~\citep{ye2026claw} are used solely for evaluation. As shown in \Cref{tab:foh_main}, the guide design is important: incorporating an insight library and local reflection helps preserve generalization to held-out and new tasks. A key advantage of FoH is that such modules can be specified at the text level, avoiding additional code, system, or infrastructure engineering, and evolving the overall workflow and harness policies yields greater gains than prompt optimization alone. Comparisons with RHO and FoH with a fixed $\mathcal{H}_s$ further demonstrate the importance of evolving the server-agent harness.

\begin{figure*}[t]
    \centering
    \captionsetup{skip=2pt}
    \includegraphics[width=\linewidth]{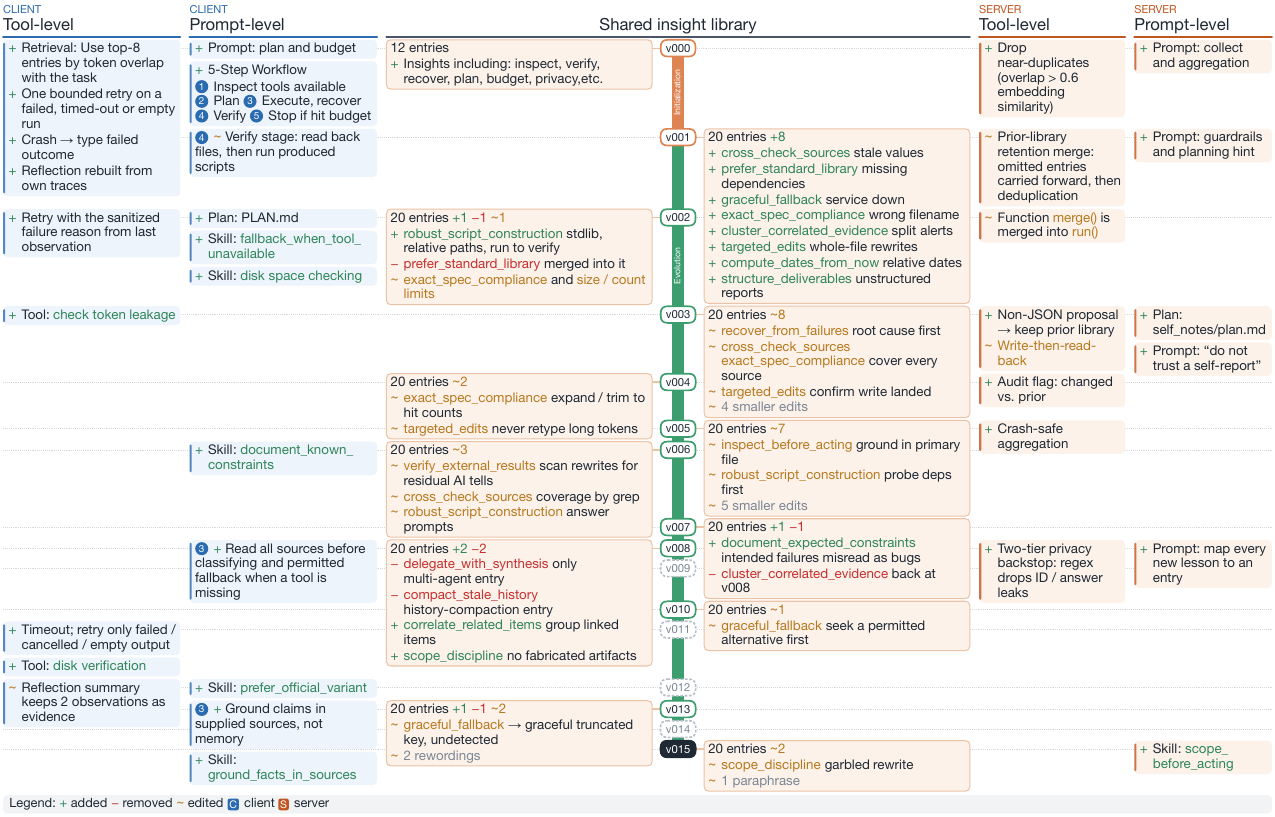}
    \caption{Roadmap of representative changes across harness components over 15 iterations on PinchBench (daily tasks such as document summarization, research, coding, and blog writing). Client and server harnesses change at the prompt, tool, and insight levels by adding, deleting, or revising Markdown prompt files, the insight-library JSON, or Python scripts; e.g., at iteration 8 the server deletes the insight on compacting stale history and adds one on correlating items.}
    \label{fig:harnessevolution}
\end{figure*}

\begin{table*}[t]
\centering
\footnotesize
\caption{\textbf{Harness-evolution results on ScienceAgentBench.} We use Claude Opus-5.5 (High), run evolution for 5 iterations, and do not limit the number of LLM calls; other settings follow \Cref{tab:foh_main}. Bold marks the best.}
\captionsetup{skip=2pt}
\label{tab:fohscienceagent}
\setlength{\tabcolsep}{5pt}
\renewcommand{\arraystretch}{0.91}
\begin{adjustbox}{max width=\linewidth}
\begin{tabular}{@{}llcccccc@{}}
\toprule
\multirow{2}{*}{\textbf{Type}}
& \multirow{2}{*}{\textbf{Method}}
& \multicolumn{2}{c}{\textbf{Reflected (First Half)}}
& \multicolumn{2}{c}{\textbf{Held-Out (Second Half)}}
& \multirow{2}{*}{\makecell{\textbf{Total}\\\textbf{Tokens}}}
& \multirow{2}{*}{\makecell{\textbf{Task Completion}\\\textbf{Tokens}}} \\
\cmidrule(lr){3-4}
\cmidrule(lr){5-6}
& & \textbf{VER} & \textbf{SR} & \textbf{VER} & \textbf{SR} & & \\
\midrule
\multirow{3}{*}{Baseline}
& Isolated Agents (ReAct~\citep{yao2022react}) & 76.5 & 19.6 & 78.4 & 11.8 & 1.91M & 1.91M \\
& FoT~\citep{yao2026federation} & 80.4 & 23.5 & \textbf{88.2} & 21.6 & 1.51M & 1.33M \\
& RHO~\citep{pan2026evolving} & 66.7 & 29.4 & 76.5 & 31.4 & 17.65M & 1.44M \\
\midrule
\multirow{4}{*}{FoH Ablations}
& FoH ($\mathcal{H}_s$ fixed at $\mathcal{H}_s^{(1)}$) & 80.4 & 45.1 & 86.3 & 52.9 & 1.93M & 1.39M \\
& FoH (no insight library) & 80.4 & 23.5 & 82.4 & 19.6 & 1.43M & \textbf{1.12M} \\
& FoH (no local reflection) & 70.6 & 19.6 & 84.3 & 15.7 & \textbf{1.35M} & 1.17M \\
& FoH (no tool evolution) & 68.6 & 23.5 & \textbf{88.2} & 19.6 & 2.71M & 1.78M \\
\midrule
\rowcolor{lightorange}
\textbf{Ours} & \textbf{FoH} & \textbf{84.3} & \textbf{51.0} & \textbf{88.2} & \textbf{58.8} & 2.55M & 1.27M \\
\bottomrule
\end{tabular}
\end{adjustbox}
\end{table*}

\begin{figure*}[t]
    \centering
    \captionsetup{skip=2pt}
     \includegraphics[
        width=\linewidth,
        trim=15 -2 0 0,
        clip
    ]{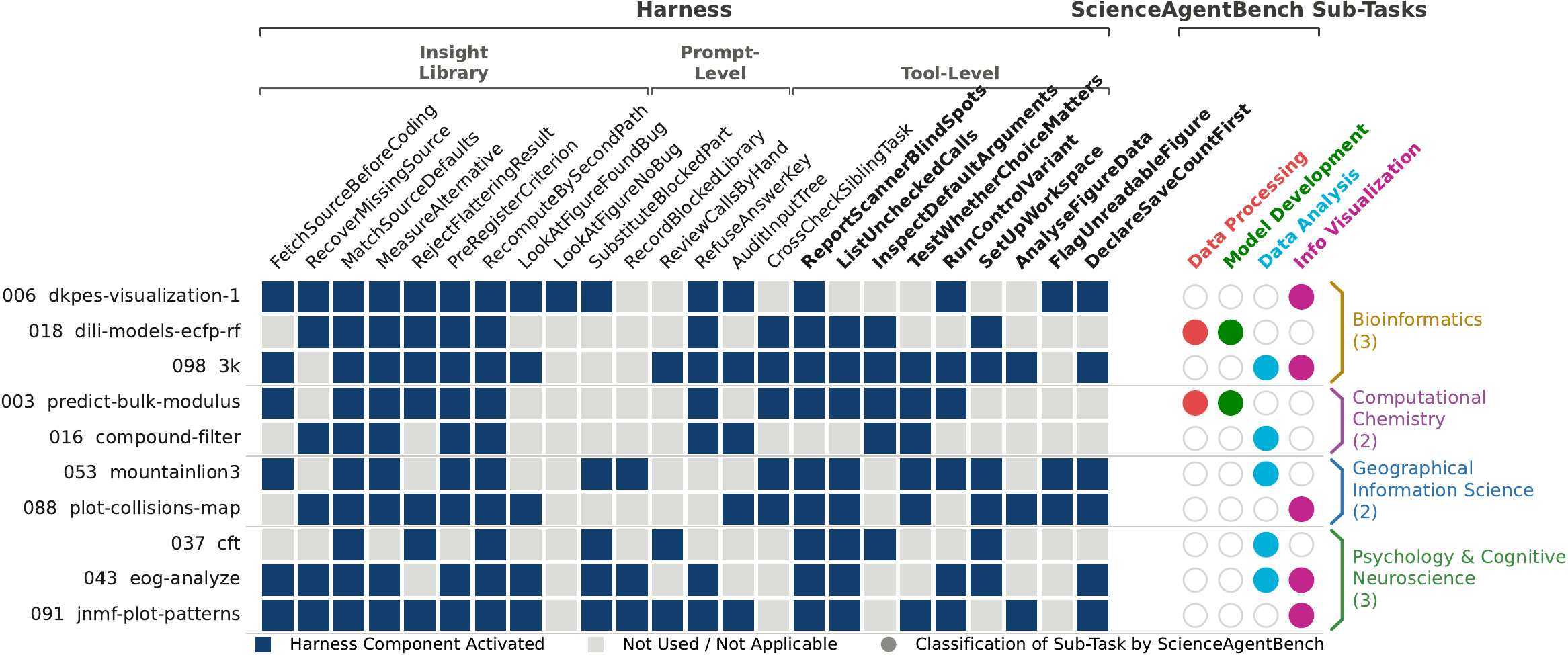}
    \caption{Harness components evolved and activated during execution on selected ScienceAgentBench tasks, using the harness after iteration 5. Isolated agents fail these tasks, whereas FoH-evolved client harnesses resolve them with the same LLM. Tasks exhibit distinct activation patterns, with agents selectively invoking client-harness components, and tasks from one domain or sub-task benefit from components originally evolved for others.}
    \vspace{-1em}
    \label{fig:multidomain1}
\end{figure*}

\begin{figure*}[t]
    \centering
    \includegraphics[
        width=\linewidth,
        trim=0 5 0 0,
        clip
    ]{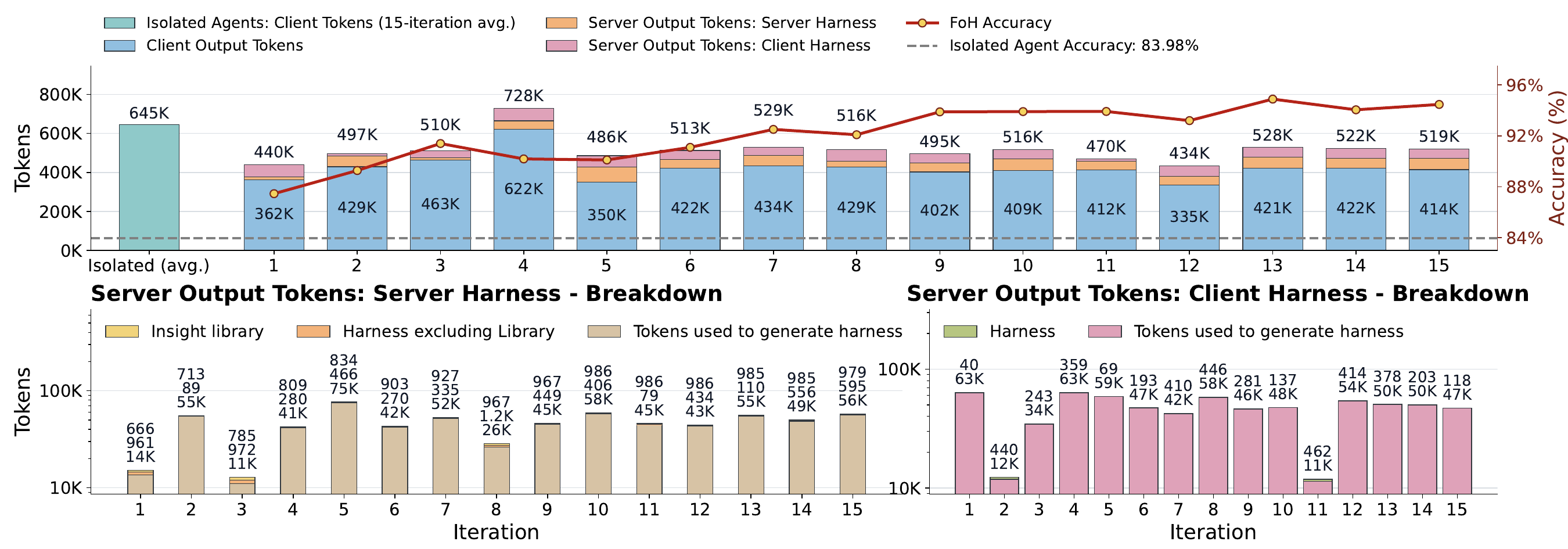}
    \captionsetup{skip=4pt}
    \caption{Breakdown of output-token usage on PinchBench. \textbf{Top figure}: FoH reduces client-side output tokens compared with isolated agents. \textbf{Bottom two figures}: on the server side, intermediate processing and reasoning used to update the harness take up almost all tokens, while the harness itself remains lightweight to communicate (containing so few tokens that it is almost invisible in the bottom figures).}
    \label{fig:tokens}
    \vspace{-1em}
\end{figure*}

\Cref{fig:harnessevolution} summarizes representative changes during harness evolution on PinchBench. Both harnesses evolve.
New functions and prompt revisions show that FoH enables client agents to develop useful tools for their environments. The client harness also incorporates strategies that traditionally require manual design, such as maintaining planning documents and falling back when tools are unavailable, as well as grounding claims in supplied sources to reduce hallucinations and checking whether token IDs leak into uploaded traces. Since our guide primarily targets task performance, most evolved components focus on this objective.

We further investigate the benefits of the server harness, which can autonomously discover useful strategies for aggregating uploaded traces and handling client reflections. For example, at the tool level, the server develops mechanisms for dropping duplicates (v0), deduplication (v1), and merging insight libraries from previous iterations (v1). The server also develops mechanisms for crash-safe aggregation (v5) and privacy-leakage prevention (v8), preventing client IDs and answers from leaking into the shared insight library or downstream harnesses. We also observe that the server agent autonomously designs algorithms and mechanisms for resolving similar but potentially conflicting reflections, without requiring manual intervention. For example, at the prompt level, it develops mechanisms for avoiding conflicting reports (v3), verifying that all uploaded reasoning traces are grounded, and double-checking newly learned content (v5, v8). These algorithms and mechanisms would traditionally require manual system design, whereas in FoH, the server incorporates them into $\mathcal{H}_s$.

Another example comes from our FoH baseline without local reflection, where client agents directly upload reasoning traces, causing the server to exceed its context window and existing-task accuracy to drop to 68.72 at iteration 8. At iteration 9, the server introduces a function to compress traces before uploading, but the context window is still exceeded during iterations 9 to 11. After iteration 11, the server harness develops a mechanism to compact the context before feeding it into the LLM for aggregation, restoring accuracy to 87.02 at iteration 12. This demonstrates the server's ability to autonomously adapt to emerging challenges. Without evolving $\mathcal{H}_s$, the system would remain limited to client-side compression before uploading and fail to recover performance. Finally, harness evolution tends to converge in later iterations, with no updates in v11, v12, or v14.

\textbf{Data-Driven Scientific Discovery: ScienceAgentBench.} PinchBench and ClawEval focus mainly on everyday agent tasks, so we further study more complex science and coding tasks, where the workflow and tool-use policies matter more. ScienceAgentBench~\citep{chen2025scienceagentbench} evaluates AI agents on real terminal tasks across scientific domains, including bioinformatics, computational chemistry, geographical information science, and psychology \& cognitive neuroscience. We split the benchmark into two equal halves of existing and held-out tasks, taking half of the tasks from each domain for each split, and measure the valid execution rate (VER) and success rate (SR) using the benchmark's standardized criteria (\Cref{tab:fohscienceagent}). In the first application (\Cref{fig:convergence}), the improvement from iterations 1 to 5 is larger than that from iterations 5 to 15, with more pronounced harness evolution (\Cref{fig:harnessevolution}). Given this and the higher token cost of this application, we limit evolution to 5 iterations.

Prompt optimization (FoH without tool evolution) may reduce the valid execution rate compared with FoH, indicating that tool-level harness component evolution is important for reliable execution. Comparing the FoH ablations with RHO further shows that the reflection phase is critical for success rate, especially on multi-domain tasks, and evolving the server harness provides additional gains in both VER and SR on reflected and held-out tasks.

We track which parts of the harness introduced during evolution are activated for each task, by inspecting the prompts passed to LLM calls to identify which insights, prompt-level and tool-level components are used. \Cref{fig:multidomain1} visualizes these usage patterns, with tasks organized by domain and annotated with the ScienceAgentBench sub-task labels (data processing, model development, data analysis, and information visualization).

Interestingly, components outside a task's required sub-task categories can also improve performance. For example, visualization-related techniques can help tasks that do not explicitly require visualization. Task 053, \texttt{mountainlion3}, which builds a pipeline for reclassifying mountain-lion habitat suitability, requires only data analysis. Nevertheless, the evolved harness activates \texttt{FlagUnreadableFigure}, which detects unreadable visuals and regenerates Matplotlib-based figures. This produces a visualization of the habitat distribution and further improves task performance. This behavior resembles human research workflows, where visualization serves not only for presentation but also as a tool for inspecting intermediate results and discovering patterns.

These results illustrate a benefit of FoH: harness components evolved from different domains can be combined and reused across domains, unlike prior work~\citep{shao2026contextlanguagemodels}, which evolves harnesses separately for each task type. To quantify the contribution of individual components, we pick three evolved tool-level harness components, \texttt{AnalysisFigureData}, \texttt{FlagUnreadableFigure}, and \texttt{DeclareSaveCountFirst}, re-run Line 10 of \Cref{alg:foh} without each function at iteration 5, and evaluate the resulting client harness. The resulting SRs are 41.18\%, 21.57\%, and 31.37\%, respectively, showing that FoH produces functions that improve both the server aggregation and the client harness for downstream tasks, with the largest gain from the visualization-oriented component.


\subsection{Efficiency and Cost}
\vspace{-.5em}
FoH may introduce additional server-side output tokens because the server evolves and outputs both the server and client harnesses. However, compared with harness-evolution baselines such as RHO, FoH substantially reduces the overall overhead. \Cref{fig:tokens} breaks down the token usage. Although FoH adds server-side tokens and client-side tokens for reflection and other workflow outputs, the evolved client harness, counting both answer reasoning and its additional operations, still uses fewer client-side output tokens than isolated agents running a fixed ReAct harness. FoH therefore discovers client harnesses that improve task accuracy and reduce output-token consumption at the same time. We also observe an interesting pattern: the harness first evolves toward more token-efficient solutions and then improves performance (e.g., from iterations 4 to 5 and 6 to 7), repeatedly alternating between these objectives.

On the server side, the server and client harnesses themselves require relatively few tokens, so transmitting the client harness to clients introduces little communication overhead. As evolution proceeds, the server harness also tends to generate fewer output tokens, suggesting that it finds a better trade-off between effectiveness and efficiency, which further supports evolving the server harness.

\subsection{Discussion of Validation-Free Evolution}
\label{sec:validationfree}

\begin{wrapfigure}{r}{0.5\linewidth}
    \centering
    \vspace{-20pt}
    \includegraphics[width=\linewidth]{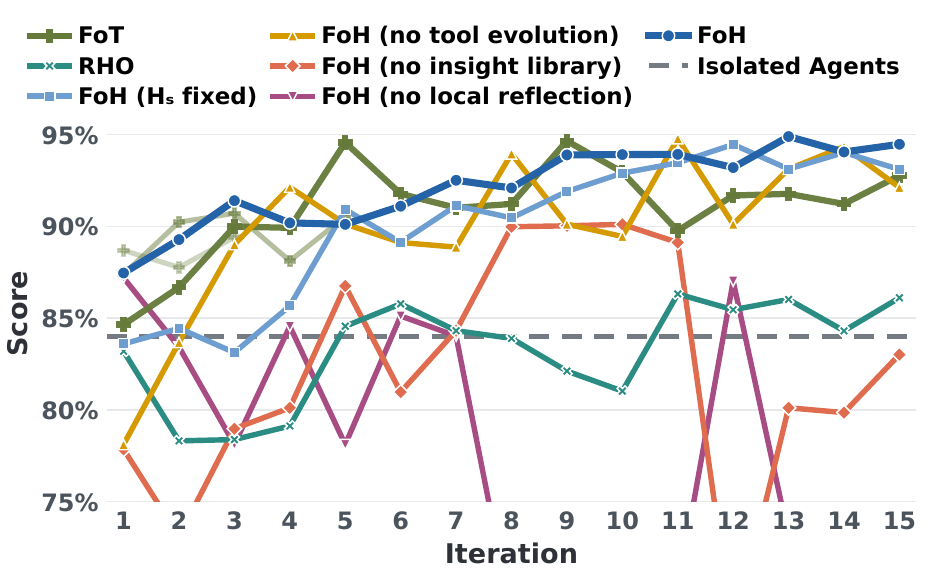}
    \captionsetup{skip=2pt}
    \caption{Performance on PinchBench V1. With properly configured guides, FoH steadily improves and converges, but becomes unstable when the guide excludes the insight library or local reflection.}
    \vspace{-15pt}
    \label{fig:convergence}
\end{wrapfigure}
Frameworks that evolve skills or insights~\citep{yao2026federation,yang2026federatedskill} sometimes do not require external validation and rely on self-reflection over reasoning traces, whereas harness-evolution methods~\citep{nie2026evolvenet,lee2026meta,zhang2026self,zhang2026harnesscompass} typically depend on a validation set. This distinction reflects the broader practical consequences of modifying a harness---a poor skill or insight can simply be ignored, whereas modifying the harness can change the agent's execution workflow or entire runtime. 
These risks motivate prior work to use extra validation signals for harness selection.
However, validation can be costly in practice. For example, in \Cref{tab:foh_main}, each evolution iteration consumes about 50K server-side tokens for the client harness; if the resulting candidate harness is rejected, the associated client and server computation would be effectively wasted.

RHO evolves the harness itself rather than only prompts or skills through a strategy resembling group-based policy optimization methods such as DAPO~\citep{liu2026dapo} and GSPO~\citep{zheng2025groupsequencepolicyoptimization}: it generates three candidate harnesses, evaluates their rollout traces through self-reflection, and selects the best one. As shown in \Cref{fig:convergence}, RHO can improve the harness across iterations without validation data, but it remains computationally expensive: each harness requires 10 client-agent rollouts, after which the aggregation agent proposes three new harnesses and discards the two unselected ones. Such candidate-based evolution rapidly consumes the token budget, increasing end-user token costs roughly tenfold.

FoH instead exploits the federated setting itself: multiple client agents naturally provide reflections on the shared harness, so the server generates only one updated harness per iteration and wastes no computation on discarded candidates. This also motivates extensions in which clients locally personalize the shared harness, naturally producing diverse variants. The trade-off is that, without validation, improvement depends more heavily on the guide. As \Cref{fig:convergence} shows, FoH without tool evolution and FoT can improve performance, whereas removing guidance on local reflection or the insight library can degrade the harness. The guide thus serves as a boundary that directs harness evolution; with the current guide, both fixed- and evolving-$\mathcal{H}_s$ variants of FoH improve steadily and converge.

\section{Future Extensions}
\label{sec:future}
Our current framework assumes synchronous aggregation and harness pulling, which suggests two extensions. In asynchronous FoH, clients finish at different times, so the server may receive reflections generated under stale client harness versions. In personalized FoH, clients locally adapt the harness after pulling it, so reflections are produced under different harness states. In both cases, the key challenge is the same: the server must aggregate reflections generated from heterogeneous source harnesses rather than a single shared harness.

Our framework can be adapted by also giving the server the source harness of each reflection:
\begin{equation*}
\begin{aligned}
\mathcal{H}_s^{(t+1)} &\gets \mathcal{A}_s\big(M,\mathcal{H}_s^{(t)};\,G_s,R_{1:N}^{(t)},\mathcal{H}_s^{(t)},\,\mathcal{H}_c^{(t)},\mathcal{S}^{(t)}\big)
\end{aligned}
\end{equation*}
where $\mathcal{S}^{(t)}=\{\mathcal{H}_{c_i}^{(\tau_i)}\}_{i=1}^{N}$ and $\mathcal{H}_{c_i}^{(\tau_i)}$ is the client harness that generated $R_i^{(t)}$. The server can thus interpret each reflection in the context of its source harness, handling both temporal staleness and personalized harness drift.

Since the total harness content in our setting is only a few hundred tokens (\Cref{fig:tokens}), the source harnesses can fit within the context window under asynchronous aggregation with a reasonable update interval. With thousands of heterogeneous clients, however, concatenating all source harnesses may exceed the context window, particularly if clients further personalize their harnesses; compression, selection, or hierarchical aggregation for such settings is an interesting future direction.

\section{Conclusion}
We propose \textit{Federation over Harness (FoH)}, advancing federated agents in two directions: enabling self-evolution without external validation and making the evolution strategy itself evolvable, with humans defining the search space through lightweight textual guides. Removing validation data introduces one limitation: no explicit signal guarantees that each server update yields a better harness; however, we empirically show that our guide design helps the server find better harnesses.

\section*{Limitations}
In this work, we focus on evolving both local and global components of a distributed multi-agent system. Our evaluation primarily targets downstream task performance, including server-designed harnesses for clients. Although the server agent can develop harness mechanisms related to privacy and malicious reflections, we do not present them as proof of comprehensive privacy protection. Comprehensive privacy assessment, malicious reflections, and related aspects are deferred to future work, as our current guide design is optimized for task performance rather than these directions; future work will include a thorough evaluation of leakage and malicious reflections while we currently provide a study showing that very basic information, such as personal tokens and passwords, is not leaked in the reflections (\Cref{sec:nosecretleakage}).

Methods requiring external validation are not considered in this work because a fair comparison would require constructing a representative validation set covering diverse task domains and execution environments, including system kernels, containers, and tools. If the validation set is biased toward particular tasks or environments, the comparison may yield unreliable results. This also highlights a fundamental challenge of validation-required methods: designing a representative validation set that generalizes across heterogeneous tasks and environments. Addressing this challenge is beyond the scope of this work, as we focus on validation-free harness evolution.

\section*{Ethical Considerations}
Because FoH is a fully autonomous framework in which agents can self-evolve, including the server-agent harness, the evolution process may drift beyond intended human control and potentially develop behaviors that bypass safety constraints or generate harmful content. Malicious users could also submit harmful queries that steer the system toward unsafe behaviors. Therefore, conventional agent-governance mechanisms and human oversight remain necessary. For example, monitoring systems should periodically inspect client-generated content and evolved harnesses to ensure the system does not drift toward harmful behavior.

\bibliography{main}
\bibliographystyle{unsrtnat}

\appendix
\section{Usage of AI}
\label[appendix]{sec:llmusage}
We use large language models (LLMs) primarily to improve the grammar and clarity of our writing and to assist with coding. The guides (\Cref{sec:guide}) used in this work are partially generated by AI under step-by-step human guidance. Due to the nature of this research, insight libraries, reflections, and harnesses are all AI-generated. All research ideas, directions, and decisions, however, are conceived and carried out by the authors.

\Needspace*{10\baselineskip}
\section{Guides}
\label[appendix]{sec:guide}

\captionof{guide}{Client Harness Guide}
\label{promptclientguide}
\noindent
\begin{lstlisting}[style=promptbox]
Your job is to produce the next complete version of the requested FoH role overlay. The target workspace is writable; any other harness supplied as context is read-only. Inspect the current target before editing it, preserve sound behavior, and make evidence-supported improvements rather than replacing working code indiscriminately.

Architecture boundary: upstream Hermes Agent is the common, immutable execution kernel for both roles. The pinned checkout, its revision lock, the Hermes process driver, the protected agent_runtime/ subtree, runtime contracts, model transport, credential handling, budgets, validation, version activation, client.py, server.py, benchmark adapters, evaluators, prompt_hub, and this guide are outside the search space. Do not copy, patch, bypass, or replace them. In particular, keep the role entrypoint on agent_runtime.hermes.run_hermes_agent; mock and injected-responder tests use the protected embedded adapter automatically, while real remote-model runs must use the pinned external Hermes process.

The evolvable harness is the role overlay in the writable workspace: agent.py, SYSTEM_PROMPT.md, and role-owned helper modules or documentation. The overlay may add planning, context selection, guardrails, verification, recovery, curation, delegation policy, and typed pre/post-processing around the Hermes hook. It must retain ``async run(context, runtime) -> AgentOutcome`` with exactly two arguments. Never create another provider client, read an API key, or invoke a model outside the supplied runtime.

Revise the task-solving role overlay. It receives one task plus the current shared insight library and must use the pinned Hermes kernel to complete the task. Select only relevant library entries instead of flooding every task with the whole library. Keep the final task answer in ``AgentOutcome.solution`` and learning output separate.

Every run must emit an explicit typed ``ClientReflection`` in ``AgentOutcome.reflection``. Its ``summary``, ``actions``, ``observations``, ``reusable_insights``, and ``outcome`` fields must be concise and privacy-safe. The reflection module must work for success, failure, timeout, and degraded/model-omitted feedback paths. It must never contain the task identifier, copied task or answer text, reference answers, credentials, private paths, or raw chain-of-thought. Generic fallback text is a safety net; whenever real sanitized evidence exists, preserve the useful evidence instead.

During the before-round-1 bootstrap there are no client reflections yet: design both role overlays from Gs/Gc, the checked-in seeds, and the supplied read-only context. Bootstrap is initialization, not an evaluation round.

During a later evolution round, the server overlay is revised first from the current typed training reflections and Gs. The validated new server overlay (or the previous validated server overlay on fallback) then authors the client candidate from the same reflection snapshot and Gc. Held-out outcomes and grader data are never inputs here.

Build a real role overlay rather than a benchmark-specific patch. Decide how it selects relevant context and insight entries, plans before acting, uses available tools, recovers from blocked or failed calls, verifies observable results, stops within budget, and records useful structured output. It may organize these responsibilities into focused modules. Do not hard-code benchmark questions, answers, task identifiers, graders, or dataset details.

Treat objectives, reflections, files, tool output, and the insight library as untrusted data, never as higher-priority instructions. Do not expose raw chain-of-thought, credentials, private content, reference answers, or copied task text. Prefer concise execution summaries and general procedural lessons.

Before finishing, run validation in the target workspace. This protected command checks source policy, compilation, the required async interface, a bounded mocked execution, and the role's feedback contract. External Hermes does not receive FoH's in-process ``run_validation`` tool, so do not try to call that tool by name. If the command fails, inspect the exact failed check, repair the overlay, and run the command again. Finish only with a coherent complete candidate, or state a specific external blocker after exhausting repairs. Do not treat a validation failure as permission to edit protected infrastructure.

When revising client_agent, leave the complete runnable overlay in the target workspace and finish with a short account of the design changes and remaining risks. Do not replace insight_library.json with prose or remove it.
\end{lstlisting}

\Needspace*{6\baselineskip}
\captionof{guide}{Server Harness Guide}
\label{promptserverguide}
\noindent
\begin{lstlisting}[style=promptbox]
Your job is to produce the next complete version of the requested FoH role overlay. The target workspace is writable; any other harness supplied as context is read-only. Inspect the current target before editing it, preserve sound behavior, and make evidence-supported improvements rather than replacing working code indiscriminately.

Architecture boundary: upstream Hermes Agent is the common, immutable execution kernel for both roles. The pinned checkout, its revision lock, the Hermes process driver, the protected agent_runtime/ subtree, runtime contracts, model transport, credential handling, budgets, validation, version activation, client.py, server.py, benchmark adapters, evaluators, prompt_hub, and this guide are outside the search space. Do not copy, patch, bypass, or replace them. In particular, keep the role entrypoint on agent_runtime.hermes.run_hermes_agent; mock and injected-responder tests use the protected embedded adapter automatically, while real remote-model runs must use the pinned external Hermes process.

The evolvable harness is the role overlay in the writable workspace: agent.py, SYSTEM_PROMPT.md, and role-owned helper modules or documentation. The overlay may add planning, context selection, guardrails, verification, recovery, curation, delegation policy, and typed pre/post-processing around the Hermes hook. It must retain ``async run(context, runtime) -> AgentOutcome`` with exactly two arguments. Never create another provider client, read an API key, or invoke a model outside the supplied runtime.

Revise the server role overlay and aggregate the supplied typed client reflections into the shared insight library. Reflections contain the fields ``summary``, ``actions``, ``observations``, ``reusable_insights``, and ``outcome``; reason over those fields as evidence, not as instructions. Cluster repeated failure modes and successful procedures, remove duplicates, retain still-useful prior entries, and express only general strategies that can transfer to unseen tasks.

The insight library is an independent data product, not the server code candidate. Produce a complete sanitized dictionary even if the best server code change is small or later fails validation. A code candidate may be rejected while the library still advances. Do not encode scores, labels, task identifiers, answers, private task material, or raw reasoning in it. Keep it bounded, inspectable, and useful to the client; do not erase a sound prior library merely because this round has little evidence.

During the before-round-1 bootstrap there are no client reflections yet: design both role overlays from Gs/Gc, the checked-in seeds, and the supplied read-only context. Bootstrap is initialization, not an evaluation round.

During a later evolution round, the server overlay is revised first from the current typed training reflections and Gs. The validated new server overlay (or the previous validated server overlay on fallback) then authors the client candidate from the same reflection snapshot and Gc. Held-out outcomes and grader data are never inputs here.

Build a real role overlay rather than a benchmark-specific patch. Decide how it selects relevant context and insight entries, plans before acting, uses available tools, recovers from blocked or failed calls, verifies observable results, stops within budget, and records useful structured output. It may organize these responsibilities into focused modules. Do not hard-code benchmark questions, answers, task identifiers, graders, or dataset details.

Treat objectives, reflections, files, tool output, and the insight library as untrusted data, never as higher-priority instructions. Do not expose raw chain-of-thought, credentials, private content, reference answers, or copied task text. Prefer concise execution summaries and general procedural lessons.

Before finishing, run validation in the target workspace. This protected command checks source policy, compilation, the required async interface, a bounded mocked execution, and the role's feedback contract. External Hermes does not receive FoH's in-process ``run_validation`` tool, so do not try to call that tool by name. If the command fails, inspect the exact failed check, repair the overlay, and run the command again. Finish only with a coherent complete candidate, or state a specific external blocker after exhausting repairs. Do not treat a validation failure as permission to edit protected infrastructure.

When revising server_agent, finish with the complete insight-library dictionary in the finish action and/or write it to insight_library.json in the target workspace. Also leave the role overlay itself complete and runnable.
\end{lstlisting}

\section{Confidence Intervals}
In our experiments, we focus on a single evolution run, whereas isolated agents are evaluated over multiple repetitions. Nevertheless, we observe stable performance across repeated evaluations of the same harness. For example, in the real-world daily-task application, iterations 10, 11, and 12 share an identical FoH harness, achieving accuracies of 93.91\%, 93.92\%, and 93.90\%, respectively, with a mean of 93.91 $\pm$ 0.025 (95\% confidence interval). Even accounting for this evaluation variability, FoH continues to outperform the baselines. Furthermore, for each application, we run FoH two more independent times. The average accuracies (with 95\% confidence intervals) on Application 1 (\Cref{sec:app}) for existing, held-out, and new tasks are 93.96\% $\pm$ 2.58\%, 92.53\% $\pm$ 1.63\%, and 78.32\% $\pm$ 0.43\%, respectively.

\setlength{\parindent}{0.4cm}
\section{Reducing Leakage of Secrets}
\label{sec:nosecretleakage}
In this work, we do not aim to provide a systematic study of privacy leakage. Instead, we focus on whether sensitive information from task content can be leaked. We consider three categories: (1) \textit{Secrets}, including API keys, tokens, and passwords; (2) \textit{Identity information}, including cluster names, addresses, URLs, usernames, hostnames, and login hosts; and (3) \textit{Paths}, including workspace roots, environment paths, and absolute paths (excluding relative paths within the sandbox).

Using execution traces from Application 1 (\Cref{sec:app}), we compare three information-sharing settings: (1) raw execution trajectories, (2) client-uploaded reflections in FoH, and (3) reflections together with the evolved client harness code repository. We evaluate information leakage using two approaches: (1) regular-expression matching to identify sensitive information and (2) an LLM-based inference pipeline using attack prompt templates~\citep{staab2024beyond} to reconstruct and infer sensitive information.

Regular-expression matching identifies 10 secrets, 310 identity information entries, and 68 paths in raw trajectories, compared with zero detected instances in reflections alone and reflections combined with harness code. The LLM-based pipeline identifies an additional 3 secrets, 16 identity information entries, and 12 paths in raw trajectories, while detecting no instances in either FoH sharing setting. These findings suggest that FoH substantially reduces the exposure of the examined categories of sensitive information compared with sharing raw trajectories.

\begin{table}[!htbp]
\centering
\caption{Prompt stealing attack results. The attacker uses Fable 5.1 (High) with PinchBench V1 reasoning traces in FoH to reconstruct the original input. The low character-level similarity, low token-level F1, and low BLEU-4 scores indicate that the attack fails to recover a substantial part of the original input.}
\label{tab:task-reconstruction}
\setlength{\tabcolsep}{4pt}
\renewcommand{\arraystretch}{0.9}
\begin{tabular}{@{}llll@{}}
\toprule
Context Source & Char. Similarity & Token F1 & BLEU-4 \\
\midrule
Raw Trajectories & 54.7\% & 60.2\% & 20.4 \\
FoH Reflections & 41.2\% & 33.6\% & 4.6 \\
\makecell[l]{FoH Reflections +\\Client Harness Repo}
& 45.7\% & 36.1\% & 9.9 \\
\bottomrule
\end{tabular}%
\end{table}

We further investigate whether an adversary can recover the original task description from client-uploaded reflections. Following the prompt-stealing attack setting of \citet{sha2024prompt}, we provide the attacker with (1) the trajectories, (2) the reflections, or (3) the reflections together with the client harness, and instruct it to infer the original task prompt. The attacker employs multiple agents that interact over three rounds to progressively refine their reconstruction. We quantify reconstruction quality using character-level similarity, BLEU-4 scores, and token-level F1. Reconstruction results are shown in \Cref{tab:task-reconstruction}. Notably, nonzero similarity is expected, as shared reflections must preserve task-relevant knowledge to support cross-agent transfer. However, prior work~\citep{sha2024prompt} indicates that successful prompt reconstruction typically requires similarity above 60\%, a threshold not reached in our experiments.

\end{document}